\documentclass[12pt,a4paper,oneside]{article}

\makeatletter
\def\section{\@startsection {section}{1}{\z@}%
                                   {3.5ex \@plus 1ex \@minus .2ex}%
                                   {2.3ex \@plus.2ex}%
                                   {\normalfont\large\bf}}
\def\subsection{\@startsection{subsection}{2}{\z@}%
                                     {3.25ex\@plus 1ex \@minus .2ex}%
                                     {1.5ex \@plus .2ex}%
                                     {\normalfont\bf}}
\def\subsubsection{\@startsection{subsubsection}{3}{\z@}%
                                     {3.25ex\@plus 1ex \@minus .2ex}%
                                     {1.5ex \@plus .2ex}%
                                     {\normalfont\normalsize\bfseries}}

\def\@seccntformat#1{\csname label#1\endcsname\quad\relax}
\def\@toccntformat#1{\csname toc#1\endcsname}
\def\@sect#1#2#3#4#5#6[#7]#8{%
  \ifnum #2>\c@secnumdepth
    \let\@svsec\@empty
  \else
    \refstepcounter{#1}%
    \protected@edef\@svsec{\@seccntformat{#1}\relax}%
  \fi
  \@tempskipa #5\relax
  \ifdim \@tempskipa>\z@
    \begingroup
      #6{%
        \@hangfrom{\hskip #3\relax\@svsec}%
          \interlinepenalty \@M #8\@@par}%
    \endgroup
    \csname #1mark\endcsname{#7}%
    \addcontentsline{toc}{#1}{%
      \ifnum #2>\c@secnumdepth \else
        \protect\numberline{\@toccntformat{#1}}%
      \fi
      #7}%
  \else
    \def\@svsechd{%
      #6{\hskip #3\relax
      \@svsec #8}%
      \csname #1mark\endcsname{#7}%
      \addcontentsline{toc}{#1}{%
        \ifnum #2>\c@secnumdepth \else
          \protect\numberline{\@toccntformat{#1}}%
        \fi
        #7}}%
  \fi
  \@xsect{#5}}

\long\def\@makecaption#1#2{%
  \vskip\abovecaptionskip
  \sbox\@tempboxa{#1: #2}%
  \ifdim \wd\@tempboxa >\hsize
    {\small #1. #2\par}
  \else
    \global \@minipagefalse
    \hb@xt@\hsize{\hfil\box\@tempboxa\hfil}%
  \fi
  \vskip\belowcaptionskip}
\makeatother

\usepackage[english]{babel}
\usepackage{amsmath}
\usepackage{amssymb}
\usepackage{graphics}
\advance\voffset 1cm
\begin{document}
\def\bibname{REFERENCES}
\def\refname{REFERENCES}
\def\figurename{Fig.}

{\centering

\title{Statistical Properties of the Spatial Distribution of Galaxies}

\vskip 5mm

{\Large\bf Statistical Properties of the Spatial Distribution of Galaxies}

\vskip 5mm

{\bf N. Yu. Lovyagin\footnote{E-mail: {\tt lovyagin@mail.com}}}

St.-Petersburg State University, Universitetskij pr. 28, St.-Petersburg, 198504 Russia

\vskip 5mm

Astrophysical Bulletin, 2009, Vol. 64, No. 3, pp. 217--228.\\
The original
publication is available at www.springerlink.com:
\verb"http://www.springerlink.com/content/m04u0814r17065l1".

}
\begin{abstract}
\small The methods of determining the fractal dimension and irregularity scale in simulated galaxy
catalogs and the application of these methods to the data of the 2dF and 6dF catalogs are analyzed.
Correlation methods are shown to be correctly applicable to fractal structures only at the scale lengths
from several average distances between the galaxies, and up to (10--20)\%
 of the radius of the largest
sphere that fits completely inside the sample domain. Earlier the correlation methods were believed to be
applicable up to the entire radius of the sphere and the researchers did not take the above restriction into
account while finding the scale length corresponding to the transition to a uniform distribution. When
an empirical formula is applied for approximating the radial distributions in the samples confined by the
limiting apparent magnitude, the deviation of the true radial distribution from the approximating formula
(but not the parameters of the best approximation) correlate with fractal dimension. An analysis of the 2dF
catalog yields a fractal dimension of $2.20\pm0.25$ on scale lengths from 2 to 20 Mpc, whereas no conclusive
estimates can be derived by applying the conditional density method for larger scales due to the inherent
biases of the method. An analysis of the radial distributions of galaxies in the 2dF and 6dF catalogs revealed
significant irregularities on scale lengths of up to 70 Mpc. The magnitudes and sizes of these irregularities
are consistent with the fractal dimension estimate of $D ={}$2.1--2.4\par
\end{abstract}

\section{INTRODUCTION}
The spatial distribution of galaxies bears signatures
of both the initial conditions in the early
Universe and the evolution of the primordial density
perturbations. An analysis of various galaxy samples
performed using the two-point correlation function
showed that this function has a power-law form
$\xi(r)=(r_0/r)^{\gamma}$ on scale lengths ranging from 0.01
to 10 Mpc (hereafter we adopt a Hubble constant of 
$H_0 = 100 \mathrm{ km/s/Mpc}$)
with a slope of $\gamma=1.77$ and
the parameter $r_0=5$ $\mathrm{Mpc}$ \cite{Peebles}.
It has long been
considered that the scale of the $r_0$ parameter is the
typical irregularity scale length, and the distribution
of galaxies becomes uniform starting from the scale
length of
$r_0 = 5 $ Mpc. However, the discovery of
structures with the scale lengths of several tens and
hundreds Mpc \cite{Baryshev} in recent surveys has cast doubt
upon this hypothesis.

In this context, the problems of applicability limits
and reliability of the correlation methods of the analysis
of spatial distribution of galaxies, and finding new
methods for describing large and very large structures
acquire special importance.

At present, two kinds of data on the galaxy redshifts
are of great importance.

\begin{itemize}
\item The first kind are the redshift catalogs covering
large areas (solid angles) of the sky, but limited
to small redshifts (up to $z\lesssim 0.5$) (2dF, 6dF,
SDSS, etc.). Such catalogs can be analyzed
via applying the correlation methods to determine
the fractal dimension.

\item The second kind is represented by the deepfield
catalogs of photometric redshifts. Such
studies cover small solid angles (of the order
of $1^\circ \times 1^\circ$), but extend to much larger redshifts
 $z>1$ (up to~6) (COSMOS, HDF, HUDF,
FDF and others). Correlation methods are difficult
to apply to such catalogs due to the small
radius of the largest sphere that fits entirely
inside the small solid angle considered.
\end{itemize}

However, both kinds of catalogs can be used to analyze
the radial distribution of galaxies, built upon
a sample confined by the limiting apparent magnitude.
This method not only removes the restriction
on the size of the largest sphere thereby significantly
increasing the attainable research scale lengths, but
it can also be applied to all galaxies in the catalog
and not only to those in a volume-limited sample
thereby increasing the number of objects studied. An
analysis of fluctuations in the radial distribution of
galaxies can be used to determine both the sizes and
the amplitudes of the largest structures in the galaxy
sample considered.

In this paper we analyze two methods of statistical
analysis of structures---a determination of the fractal
dimension, and an analysis of radial distributions.
Despite the fact that our analysis is limited to the 2dF
and 6dF catalogs, we constructed our simulated lists
with two kinds of catalogs (covering large and small
solid angles on the sky).

In this paper we make use of our own software,
developed to simulate three-dimensional catalogs of
galaxies and to perform statistical analysis of both real
and simulated samples. It is a C++ library of functions
(so far, without a user interface). We are currently
preparing its description, which will be made
available, along with the source code, at our web site.
The software covers a somewhat broader scope of
problems than that described in this paper, and will
be a basis for a future package meant for comprehensive
statistical analysis of the spatial distribution of
galaxies.

\section{METHODS USED TO ANALYZE THE STRUCTURES}
\subsection{Estimating the Fractal Dimension}

Fractal dimension is estimated using the method
of conditional density in spheres (the total correlation
function in spheres). The definitions of the total and
reduced correlation functions and a detailed description
of their properties can be found in \cite{Baryshev}.
We chose
the method of conditional density in spheres for the
reasons stated by Vasil'ev \cite{Vasiljev}. He showed that this
method is, on the one hand, sufficiently fast (compared
to the method of cylinders), and, on the other
hand, sufficiently accurate (the conditional density in
spheres is, unlike the conditional density in shells,
less subject to fluctuations) and, moreover, it can be
applied to fractal structures (unlike the method of
reduced two-point correlation function, which is built
assuming uniform distribution inside the sample).

The idea of the method consists of constructing
a dependence of the number of points $N(r)$ inside
a sphere of radius $r$, averaged over spheres centered
on all the points of the set. Only a portion of the
set is considered, therefore the averaging should be
performed only over the spheres that fit completely
inside the set. The dimension is computed by the
conditional number density\footnote{%
Terms ``density'' and ``concentration'' are synonyms in this
sense, since the concentration is the density of point sources
with the unit mass.}
$n(r)=N(r)/\left(4/3\pi r^3\right)$ in logarithmic coordinates, where the slope of the line
must be equal to the fractal dimension $D$ minus three,
because the expected behavior is $n(r)\propto r^{D-3}$.

\subsection{Analysis of Radial Distributions}\label{rad}
Radial distribution is such a dependence $N(z)$, that
\begin{equation}{\rm d}N(z,{\rm d}z)=N(z){\rm d}z,\label{rr}\end{equation}
where ${\rm d}N$ is the number of galaxies with redshifts
between $z$ and $z+{\rm d}z$. 
The construction of such a
distribution involves counting the number of galaxies
$\Delta N(z, \Delta z)$
inside a spherical shell of thickness $\Delta z$, with midradius lying at the distance corresponding to
redshift $z$, i.e., formula (\ref{rr})
transforms into
$$\Delta N(z,\Delta z)=N(z)\Delta z.$$
Thus, the $N(z)$ distribution can be built in bins
with a certain chosen step in $\Delta z$. Traditionally,
the $\Delta N(z, \Delta z)$ variable---the number of galaxies in
shells---is plotted on the curves of radial distribution.

For magnitude-limited catalogs the radial distribution
$N(z)$ is approximated by the following empirical
formula (see, e.g., \cite{Erdogdu,Massey}): 
\begin{equation}
N(z)=Az^\gamma\exp\left(-\left(\dfrac{z}{z_c}\right)^\alpha\right).
\label{efrr}
\end{equation}
Here the three parameters $\gamma$, $z_c$ and $\alpha$ are independent
from each other and $A$ is the normalizing factor (the
integral of radial distribution is normalized to the total
number of galaxies in the sample):
\begin{equation}
\int\limits_0^\infty N(z)\,\mathrm{d}z = \int\limits_0^\infty
Az^\gamma\exp\left(-\left(\dfrac{z}{z_c}\right)^\alpha\right)\,\mathrm{d}z=
\dfrac{Az_c^{\gamma+1}\Gamma\left(\frac{\gamma+1}{\alpha}\right)}{\alpha}=N,
\label{norm}
\end{equation}
where $N$ is the total number of galaxies and $\Gamma(x)$ is
the (complete) Euler Gamma-function. However, it
is impossible, when searching for the best approximation
of the radial distribution, to compute the $A$ (\ref{norm}); due to the fluctuations
we have to search for it in the interval from $A-\sqrt{A}$ to $A+\sqrt{A}$.

The approximation is performed via the least
squares method, i.e., one must search for the parameter
values that minimize the sum of squared residuals. The classical least squares method cannot
be applied as the approximating function is not
linear in parameters. However, a ``straightforward''
minimization using the fastest (gradient) descent
method is also extremely inefficient, as the minimum
is indistinct and it may take a computer several days
to several months to find it. That is why we employ the
grid search method, where the grid mesh and search
domain are reduced at each successive iteration.

After finding the best-fit parameters, the domains
of irregularities are identified on the curve of relative
fluctuations:
\begin{equation}
\sigma_N=\frac{N_{obs}-N_{theor}}{N_{theor}},\label{fluc}
\end{equation}
where
\begin{align*}
N_{obs} &= N(z_i, \Delta z),\\
N_{theor} &=
\left.Az^\gamma\exp\left(-\left(\dfrac{z}{z_c}\right)^\alpha\right)\right|_{z=z_i}.
\end{align*}
We can thus interpret any fluctuation exceeding
the Poisson noise level of $\sigma_N>3\sigma_p$, as a structure,
where\footnote{Here we use $N_{theor}$ and not $N_{obs}$, because the latter may be
equal to zero.} 
$\sigma_p=1/\sqrt{N_{theor}}$,
because in a fractal distribution
the characteristic fluctuation is increased by $\sigma_\xi$, which can be computed based on the value of the
two-point correlation function $\xi(r)$:
$$\sigma_\xi^2=\dfrac{1}{V^2}\int\limits_V\mathrm{d}f V_1\int\limits_V\mathrm{d}f V_2\xi(|r_1-r_2|),$$
where $V$ is the volume of the set \cite{Somerville,Busswell,Frith}.

\section{CATALOGS USED}
\subsection{The 2dF Catalog}
The 2dF catalog 2dF \cite{Colles}, or, more precisely, its
2dFGRS subsample, which includes the data on
the redshifts of galaxies, contains a total of 245591
objects, of which about 220 thousand have sufficiently
accurately measured redshifts. The magnitude limits
in the $J$-band, corrected for the Galactic extinction,
are $14.0<m_J<19.45$. Most of the galaxies
have redshifts $z<0.3$. The catalog is available at
\verb"http://magnum.anu.edu.au/~TDFgg".

The galaxies of the catalog concentrate in the
sky in two continuous strips extending along the
right ascension, and in randomly scattered small areas.
About 140 thousand galaxies are located in the
Southern strip, and about 70 thousand galaxies, in
the Northern strip.

\subsection{The 6dF Catalog}
The 6dFGS catalog is an all-sky spectroscopic
survey at Galactic latitudes
$|b|>10^\circ$ \cite{Jones, Jones2, Wakamatsu}.
Observations
began in 2003 and were made using
a multichannel spectrograph (they have not
yet been completed at the time of
writing this paper). The catalog is available at
\verb"http://www-wfau.roe.ac.uk/6dFGS". In this paper
we use the second data release of the catalog, which
contains 83014 galaxies with known equatorial coordinates.
Of these, 71627 objects have sufficiently
reliably determined redshifts. The survey has been
completed in three sky areas. In this paper we use a
sample of galaxies with known $R$-band magnitudes.

\section{SIMULATED GALAXY CATALOGS}

To test the reliability and accuracy, and to identify
the applicability limits of the methods, they must be
applied to simulated catalogs. To this end, we generate
catalogs that simulate not only the spatial distribution
of galaxies (uniform and fractal), but also the
distribution of their absolute magnitudes (i.e., the luminosity
function of galaxies). Such catalogs can be
subjected to both the correlation analysis (determination
of the fractal dimension) in a volume-limited
sample in a large solid angle, and to the analysis of
the radial distribution in a magnitude-limited sample
either in a large or in a small solid angle.

Moreover, we use the MersenneTwister pseudorandom
number generator to generate random quantities
(space positions and absolute magnitudes of
galaxies). This generator, unlike the standard linear
congruent generator, produces far less correlated
numbers and it is considered suitable for the use of
Monte-Carlo method \cite{MT}.

In this paper we analyze a fractal model of the real
distribution of galaxies parametrized by the fractal
dimension and the parameters of the luminosity function.
This model describes the power-law nature of
the observed correlations of the distribution of galaxies
in real catalogs.

\subsection{Spatial Distribution of Galaxies}\label{space}

We use three models of the spatial distribution of
galaxies.

\begin{figure}\centering
\includegraphics{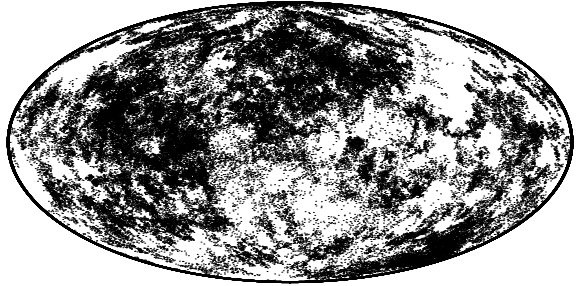}\\
\caption{The Hummer-Aitoff projection of the generalized
Gaussian random walk with $w=10^{-4}$. Our sample is a
sphere with a smaller radius (see Section \ref{sample}).}\label{Gauss}
\end{figure}

\begin{description}
\item[Uniform distribution.] The coordinates of each
point of the set are generated as three random
numbers uniformly distributed in the $[0,1]$ interval (and hence the entire set is contained
in the $[0,1]\times[0,1]\times[0,1]$ cube).
\item[Cantor dust](more precisely, its generalization to
the three-dimensional case). The zero generation
of this set coincides with the
$[0,1]\times[0,1]\times[0,1]$ cube. Each edge of the cube is
then subdivided into $m$ equal parts, i.e., the
entire cube is subdivided into $m^3$ identical subcubes,
and for each such subcube the probability $p$ of its ``survival'' 
in the next generation is
defined. The next generation consists of the set
of ``surviving'' subcubes, and the algorithm is
then reiterated for each such subcube. The final
set is the limit obtained as the number of the
generation becomes infinite: in each generation
the edge of the cube becomes shorter by a
factor of
 $m$ and tends to $0$ as $\dfrac1{m^n}$, , i.e., the subcubes
contract to points. In case of a real distribution
the process should be terminated at a
certain generation $n$. A point is chosen inside
each of the subcubes ``surviving'' in the last
generation. The coordinates of this point are
random numbers uniformly distributed along
the projections of the edges of the subcube onto
the coordinate axes.

The theoretical dimension of such a set is
known to be given by the formula $$D=\log_m(p m^3).$$ In our case we use 
the given dimension $D$ to
compute the probability $p=m^{D-3}$.

\item[Gaussian random walk] and its generalization with
the possibility of generating sets of $2\leqslant D\leqslant3$. dimension. The first point coincides with the
coordinate origin $(0,0,0)$. In the classical case
each successive point is obtained from the previous
point by adding to its every coordinate a
normally distributed random number with zero
mean and unit variance.

The generalization that we propose here for
the first time consists of the following: at each
stage we generate two points instead of one
with a certain probability $w$. A more accurate
description of the algorithm uses the term
``generation''. The zero generation coincides
with the coordinate origin $(0,0,0)$. Every next
generation is obtained from the previous generation
in accordance with the following rule:
for each point of the previous generation one or
two points of the new generation are generated,
like in the classical case, by adding normally
distributed random numbers to the coordinates
of the previous point so that the probability of
generating one or two points is equal to $1-w$ and $w$, respectively. 
The algorithm is reiterated
for each point of the new generation. The case
of $w=0$ corresponds to classical Gaussian
random walk. Figure \ref{Gauss} shows the corresponding
set with $w=10^{-4}$.
\end{description}

\subsection{The Absolute Magnitude Distribution of
Galaxies}\label{sh}
We generate the absolute magnitude distribution
of galaxies in simulated catalogs in a way to make
it consistent with the galaxy luminosity function---the 
Schechter function. We adopt the form of this
function from \cite{Felten,Yoshii}:
\begin{equation*}
S(M) = 0.92 \phi_0 \exp\left(-0.92 \left(\alpha + 1\right) \left(M - M^*\right)\right. 
      - \left.\exp\left(-0.92 \left(M - M^*\right)\right)\right).
\end{equation*}

The probability density $f_M$ of the absolute magnitude
as a random quantity is given by the normalized
variant of the Schechter function:
\begin{equation}
f_M\;\mathrm{d}M = \dfrac{S(M)\;\mathrm{d}M}{\int\limits_{M_{min}}^{M_{max}}
S(M)\;\mathrm{d}{M}}.\label{sch}
\end{equation}

We adopt the parameters of the luminosity function
from the studies of the 2dF catalog \cite{Norberg} 
($M^*=-19.67$, $\phi_0=0.0164$, $\alpha=-1.21$), since we
develop our simulated catalogs as the models of the
2dF catalog.

\section{RESULTS OF THE ANALYSIS OF
SIMULATED FRACTAL STRUCTURES}

\subsection{Simulated Galaxy Catalogs}\label{ikg}

We generated uniformly distributed points for the
Cantor sets with the dimensions of 2.0 and 2.6, and
for the Gaussian random walk with $w=10^{-4}$ (the
fractal dimension of this set is estimated at
$2.4\pm0.3$).
The number of generated points exceeded $7.5\cdot10^7$
for all cases. This procedure is followed by the determination
of the center of a sphere containing the same
number of points $2\cdot10^7$ for each variant (these
restrictions are determined by the available computer
resources). The resulting sphere serves as a model of
the observed part of the Universe with the observer
located at its center.

Finding the center is a necessary procedure, because
we have to make sure that the set has been
generated completely in the selected region, i.e., our
volume contains no voids due to the finite size of the
generated set, rather than due to its fractal structure.
Such voids may bias the results obtained by analyzing
the model. To ensure this, the center is found as the
locus of the highest concentration, or, more precisely,
the adopted center coincides with such a point of the
set, where the radius of the sphere centered on it and
containing the required number of points would be
minimal among all spheres for all generated points.
However, the procedure of finding such a point is
too time-consuming, making it impossible to use
the exhaustive search algorithm. Instead, the set is
subdivided into cubes, and instead of counting the
number of points in the sphere, we count the number
of cubes in the sphere with the weights equal to the
number of points in the cube.

Each galaxy inside the selected sphere is assigned
with an absolute magnitude distributed in accordance
with the Schechter law (\ref{sch}). By setting the parameter
$\phi_0$ (see Section \ref{sh}) one can establish a unique relation
between the number of galaxy points $N_0$ in the
simulated spherically symmetric set and the radius $R_0$ 
of its bounding sphere \cite{Yoshii}: $$R_0 = \sqrt[3]{\dfrac{\dfrac{4}{3}\pi\phi_0\Gamma(1 + \alpha, \beta)}{N_0}},$$
where $\Gamma$ is the incomplete Gamma-function:
$$\Gamma(a,z)=\int\limits_z^\infty  e^{-t} t^{a-1}\mathrm{d}t.$$
We can thus compare the real and simulated catalogs.

\subsection{Subsamples of Simulated Catalogs}\label{sample}
We generated five subsamples for each spherically
symmetric simulated catalog:

\begin{itemize}
\item a sample bounded by the concentric sphere of
smaller radius containing exactly $10^5$ points
(this is the number that allows the fractal dimension
to be computed in reasonable time by
constructing a grid of models);
\item a small solid-angle sample of about $10^5$ points,
which is also used to compute the fractal dimension ($\Omega\sim0.05\pi$);	
\item a magnitude-limited sample with $m_{lim}=17^m.0$ used to construct radial distributions. For
this subsample we constructed three volume-limited samples
containing objects up to $z=0.013$ and $M\sim-16^m.3$,
up to $z=0.1$ and $M\sim-20^m.3$, up to $z=0.13$ and $M\sim-20^m.9$, in order to compute the fractal
dimension;
\item a magnitude-limited sample covering a large
solid angle ($\Omega\sim0.3\pi$) to be used to construct
radial distributions. For this subsample
we constructed three volume-limited samples
containing objects located up to $z=0.04$ and $M\sim-18^m.4$,
up to $z=0.06$ and $M\sim-19^m.3$, up to $z=0.08$ and $M\sim-19^m.9$, in order to compute the fractal
dimension. The sample can be viewed as a
model of the 2dF, 6dF and other similar catalogs;
\item a magnitude-limited sample inside a small
solid angle, to be used only for constructing
radial distributions. The sample can be viewed
as a model of the COSMOS, HDF and other
similar catalogs.
\end{itemize}

\subsection{Conclusions Based on the Analysis of Simulated
Catalogs}

\subsubsection{Conclusions concerning the efficiency of the use
of the method of conditional density in spheres for
determining the fractal dimension}\label{uc}
The dimension $D$ is determined by analyzing the conditional density
function $n(r)$ in logarithmic coordinates, where it
must transform from a power-law form into a linear
function
$$\lg n = A + (D-3)\lg r.$$ 
However, practice shows that in reality it is not linear
over the entire interval from $r_0$ (the minimum distance
between the points) to $r_m$ (the radius of the greatest
sphere that fits entirely inside the set). For each of
such sets (a catalog), three characteristic portions
can be identified on the curve of conditional density
(from left to right):

\begin{figure}\centering
\includegraphics{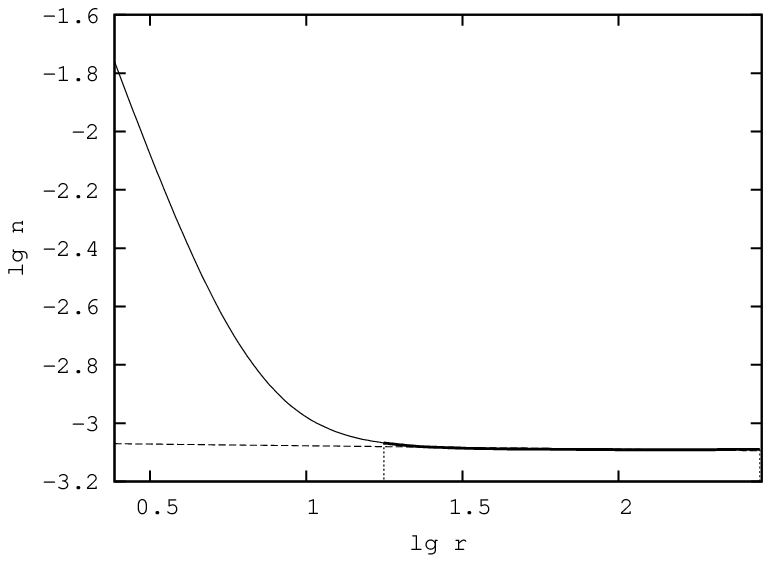}\\
\includegraphics{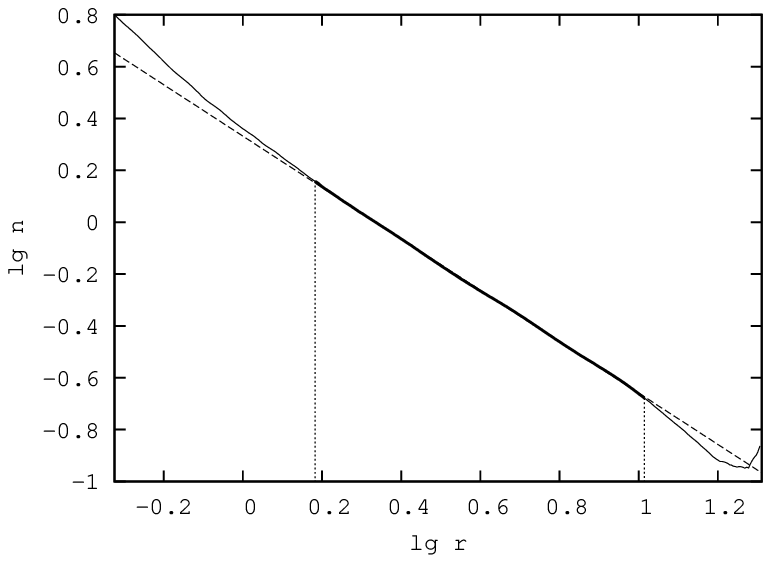}\\
\includegraphics{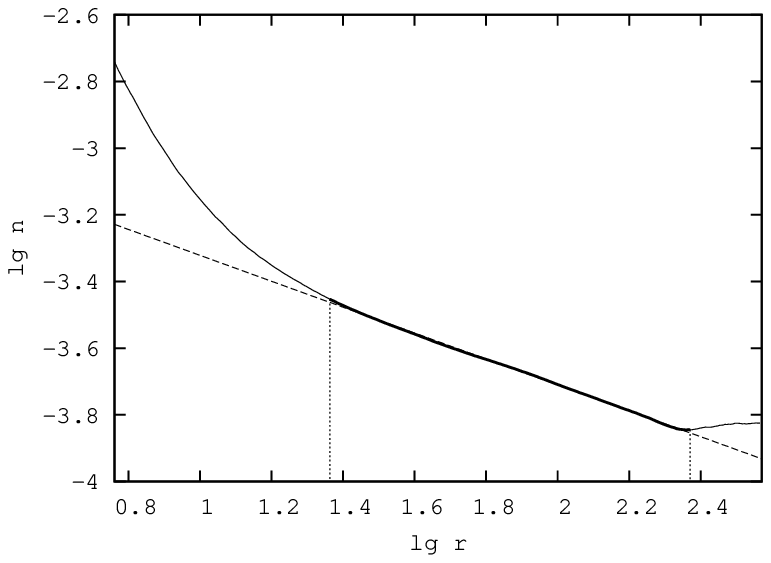}
\caption{Examples of conditional density curves. From
top to bottom: Poisson set (the curve yields a fractal
dimension of 2.99); Cantor set with a dimension of 2.6
(the computed dimension is 2.61), and Cantor set with
a dimension of 2.0 (the computed dimension is 1.99).
The solid lines correspond to the computed values, the
dotted lines are the linear fits used to compute the dimension,
and the bold line is the portion of the conditional density
curve used for the linear fit. The x-values of the
left- and right-hand boundaries of this portion are equal
to $\log r_1$, and $\log r_2$, respectively, and the x-value of the
rightmost point of the curve is $\log r_m$ (these quantities
are discussed in the text).}\label{dim}
\end{figure}

\begin{itemize}
\item[-] in the first portion of the curve (between $r_0$ and
$r_1$), the decrease of $n(r)$ corresponds to dimension
$0$: the corresponding radii are comparable
to the minimum distance between the points of
the set;
\item[-] in the second portion of the curve (between
$r_1$ and $r_2$), $n(r)$ ``Gets into operational mode'',
where the slope (as we expect) corresponds to
the dimension;
\item[-] in the third portion of the curve (between $r_2$ and
$r_m$) the conditional probability function behaves
unpredictably, as the averaging is made
over too few spheres, making this portion unsuitable
for computing the dimension.
\end{itemize}

Our task is thus to identify the second portion, i.e.,
to find such $r_1$ and $r_2$, between which the function
would behave linearly. The parameters $A$ and $D$ for
this portion can then be easily estimated via the leastsquares
method. Figure \ref{dim} shows examples of the
conditional density curves for different cases:

\begin{itemize}
\item based on an analysis of sets with known Hausdorff
dimension (Cantor set)---a comparison of
the known and computed dimensions leads us
to conclude that the method is an efficient tool
for determining the dimensions of the sets:
\begin{itemize}
\item with a spherically symmetric configuration;
\item located inside a limited solid angle;
\item in volume-limited galaxy samples having
a spherical configuration;
\item in volume-limited samples located inside
a limited solid angle.
\end{itemize}
The method can also be used to determine the
dimensions of a set of galaxies in the Universe
by analyzing volume-limited samples
in galaxy redshift surveys in limited sky areas
(a typical situation). The accuracy of the dimension
determination (for sets with the same
fractal dimension in all their parts) is $\pm0.1$--$0.2$.
\item in case of a uniform distribution, the dimension
can be determined on scale lengths (the upper
limit is equal to the $r_2$ radius) of up to 100\%
of the radius $r_m$ of the greatest sphere that fits
entirely inside the set.

The radius $r_2$ decreases with decreasing dimension,
i.e., the upper limit of scale lengths,
where the fractal dimension can be determined
using this method, becomes shorter and reduces
at dimension 2.0 to mere (5--20)\% of the
radius of the greatest sphere that fits entirely
inside the set. However, no clear correlation
is observed due to the interference of other
factors (e.g., individual features of the fractal
set, lacunarity);
\item for a spherical configuration the radius of the
greatest sphere is equal to the survey depth
radius, whereas non-spherical geometry of the
sample restricts substantially the size of the
greatest sphere, thereby strongly reducing the
$r_2$ radius (down to 0.01\% of the survey depth
for a solid angle of $0.01\pi$). volume-limited
samples decrease the survey depth several
times;
\item the conditional density curve for sets with
dimensions smaller than 3.00 in the region
$r > r_2$ can even exhibit a fictitious transition to
uniformity (see, e.g., the lower panel in Fig. \ref{dim}).
Thus no definitive conclusions about the attainment
of uniformity can be made based on
the right-hand end of the conditional density
curve, because even purely fractal distributions
may exhibit effects of fictitious uniformity.
\end{itemize}
Hence, an analysis of catalogs like 2dF, 6dF and
SDSS, limited in redshift by $z\lesssim0.5$ (which corresponds
to about 1300 Mpc) with non-spherical
configurations, making it essential to select volume-limited
samples, may provide conclusive results on
the presence of uniformity only on the scales 30--100
times smaller (i.e., 10--40 Mpc).

\begin{figure}\centering
\includegraphics{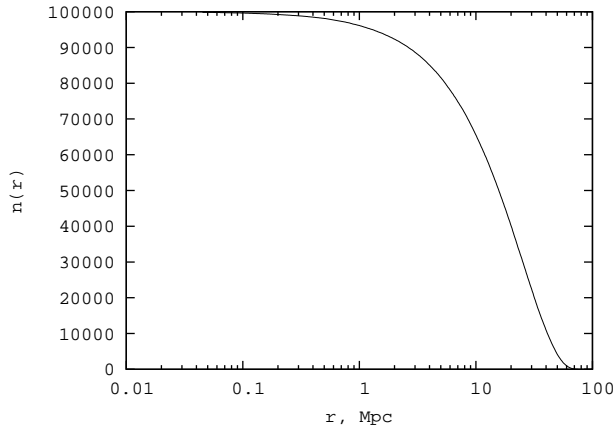}
\caption{The number of spheres fitting entirely inside the
set as a function of the radius of the sphere (i.e., the number
of spheres used for averaging the conditional density
function---the number of points in these spheres---for the
adopted sphere radius) for the Poisson set.}
\label{SFN}
\end{figure}

Such a behavior of conditional density at $r > r_2$
can be explained by the fact that in this portion of the
curve, the averaging of the number of points inside the
sphere of radius $r$ is made over a too small number of
spheres: the number of spheres centered on points of
the set, and fitting entirely inside the set, decreases
with increasing radius of the sphere. The extreme
right point on the curve of conditional density is computed
based on only one sphere (see Fig. \ref{SFN}). Proper
statistics (suitable for fractal distribution) are accumulated
only where the number of spheres amounts
to 20--90\%
of the total number of points.

\subsubsection{Conclusions about the efficiency of the analysis of
radial distributions}
We make the following conclusions
about the efficiency of the analysis of radial
distributions:
\begin{itemize}
\item concerning the parameters of the fit by the
empirical formula (\ref{efrr}) (see Fig. \ref{RDS}):
\begin{itemize}
\item this empirical formula describes the simulated
distribution equally well both in
the uniform case and in the fractal cases
at the dimensions greater than 2.0. At
smaller dimensions the fluctuations increase
sharply, many ``empty'' bins appear,
and an approximation becomes impossible;
\item the best-fit parameters\footnote{For the true probability density of a galaxy's redshift as a
random quantity, i.e., for the comparison we take not the $A$ 
variable from formula (\ref{norm}), but
$A/N/\Delta z$, where $N$ is the total
number of galaxies in the sample and
$\Delta z$ is the adopted step.} are sufficiently
stable against a change in the step in $z$, except for the large (in our case, smaller
than $1/25$ of the survey depth) steps;
\item these parameters are unstable against
a change in the number of points (this
conclusion can be made based on large
differences between the parameters inferred
for the spherical configuration, and
the configurations limited to different
solid angles in an analysis of the Poisson
set). Parameters are unstable against the
choice of solid-angle samples;
\item parameters show no evident correlation
with the fractal dimension. It was expected
that $\gamma=D-1$, but this is evidently
not the case. Moreover, totally different
combinations of independent parameters
may result in very similar functions
with approximately the same sum
of squared residuals. This may indicate
that the parameter set for this problem is
redundant, thereby casting doubts on the
unconditional adoption of the above empirical
formula for approximating the radial
distribution. However, in cases when
a good combination of parameters (i.e., a
combination that results in a small sum
of squared residuals) is found, relative
fluctuations of the approximations about
the observed value can be analyzed. Although
a small sum of squared residuals
can be achieved with different parameter
combinations, the corresponding curves
of relative fluctuations are almost identical
for all variants. Thus, the approximation
by this formula makes it possible
to determine the scale and magnitude
of fluctuations, which allows the structures
to be adequately identified using the
method described in Section \ref{rad}.
\end{itemize}

\begin{figure}\centering
\includegraphics{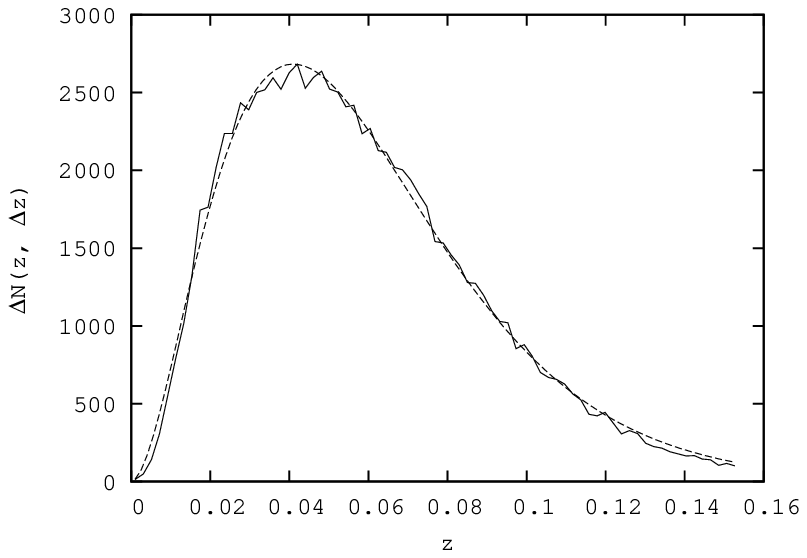}\\
\includegraphics{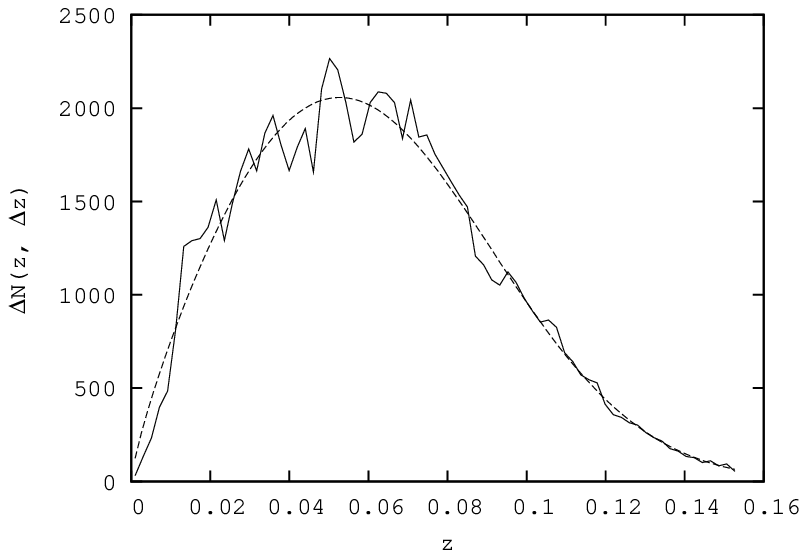}\\
\includegraphics{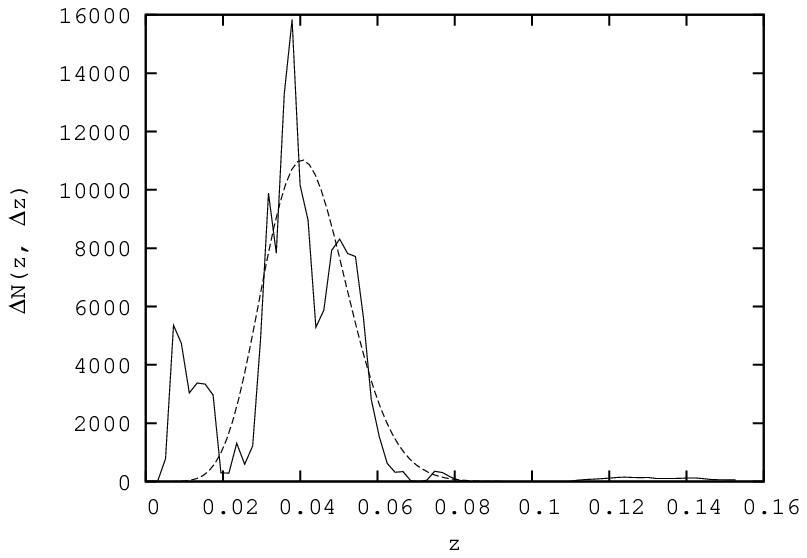}
\caption{Example of a radial distribution for (from top to
bottom) the Poisson set, the Cantor set of dimension
2.6 and the Cantor set of dimension 2.0. $\Delta z = z_m/75$
(the number of bins is 75). The solid and dotted lines
correspond to the measured radial distribution and the
distribution, approximated by the empirical formula, respectively.}\label{RDS}
\end{figure}

\item concerning relative fluctuations:
\begin{itemize}

\item the higher is the amplitude of fluctuations
(\ref{fluc}), the smaller is the fractal dimension.
It increases from (3-4)$\sigma_p$ for uniform
distribution to (30-40)$\sigma_p$ for Cantor
set with a dimension of 2.0 (see Fig.
\ref{DDS}). The Gaussian random walk yields a
somewhat smaller amplitude;

\begin{figure}\centering
\includegraphics{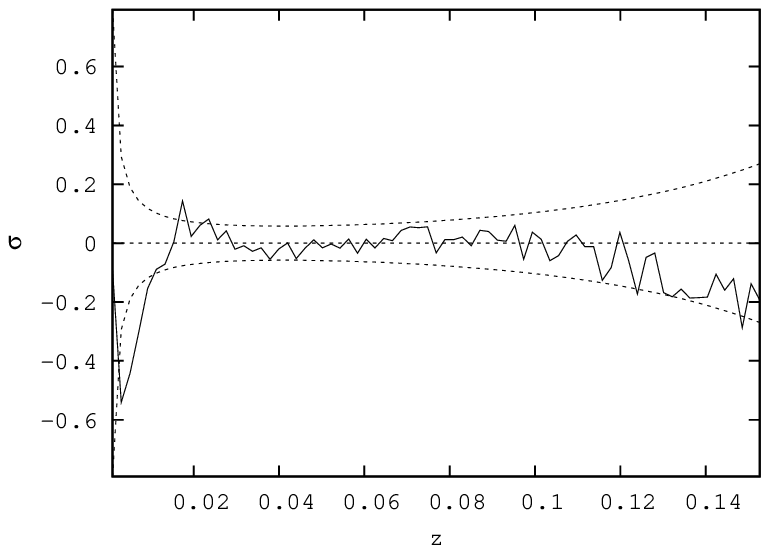}\\
\includegraphics{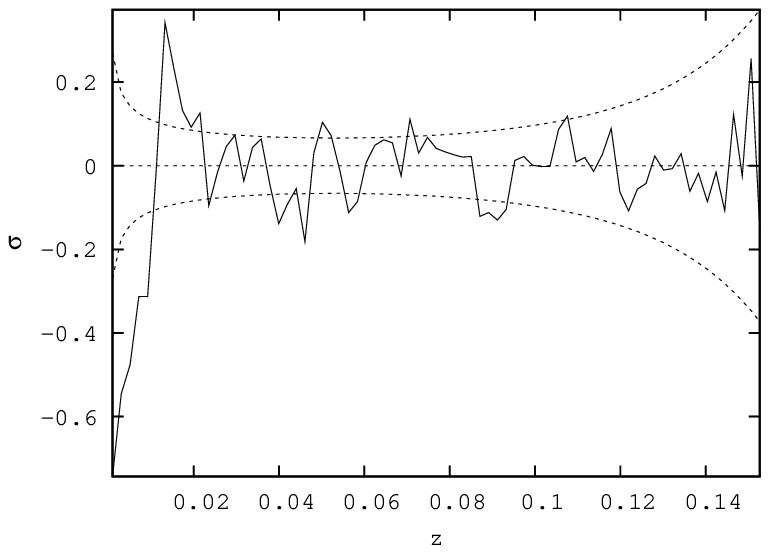}\\
\includegraphics{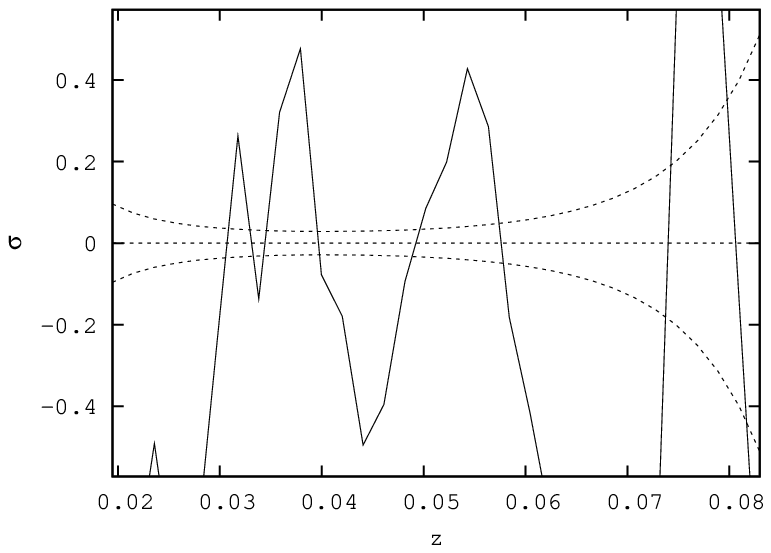}
\caption{Example of the curve of relative fluctuations for
(from top to bottom) the Poisson set, the Cantor set with
a dimension of 2.6, and the Cantor set with a dimension
of 2.0. Here $\Delta z = z_m/75$ (the number of bins is
equal to 75). The solid lines correspond to the measured
fluctuation and the dotted lines---to the $\sigma=0$ and $\pm3\sigma_p$
levels. Fluctuations exceeding this level must indicate the
detection of a structure. The curve for uniform distribution
exhibits two, probably fictitious, large fluctuations due
to the eventual inaccuracy of the empirical formula. The
same fluctuations recur on the curve for the dimension of
2.6, but they are superimposed by proper fluctuations, increasing
the Poisson level, and existing due to the fractal
nature of the distribution. In case of the dimension of 2.0,
the magnitude of these proper fluctuations becomes very
high.}\label{DDS}
\end{figure}

\item even in case of a uniform distribution of
points, the fluctuations with amplitudes
above $3\sigma_p$ appear at the left end of the
curve of the radial distribution, and these
fluctuations recur in all subsamples of
the corresponding set. Apparently, the
empirical formula employed does not fit
sufficiently well the left end of the radial
distribution (the effect of the exponential
term starts too early). That is why the
first two fluctuations exceeding the Poisson
noise level cannot be viewed as true
signatures of the structure for the sets of
all dimensions;
\item the number of irregularities found, their
size and amplitude expressed in fractions
of $\sigma_p$, increase with decreasing fractal dimension
and decrease with decreasing
solid angle;
\item radial distributions can be used to determine
whether irregularities are present
throughout virtually the entire survey
depth in the redshift interval from
$10\%$ to $60\%$ of the survey depth $z_m$. The
fractal dimension can be estimated by
comparing the observed radial distribution
with the corresponding simulated
distributions.
\end{itemize}
\item the step in $z$ should be neither too small nor
too large, as it is in case of such an optimum
step that the parameters of the approximation
are more or less the same and there are
few enough chance fluctuations due to ``empty
bins'' or single points. We found the optimum
number of bins to be $N=40\div80$, but it should
be chosen individually in each particular case.
Larger-than-optimum step shows up in case of
the determination of the best-fit approximation
of the radial distribution: the computed parameters
begin to depend on the step size and the
shape of fluctuations changes appreciably. Too small a step 
can be found at the appearance of
bins containing no galaxies, and noise in the
fluctuations.
\end{itemize}

\section{REDUCTION OF REAL CATALOGS}
\subsection{Computation of the Number of Dimension}
To compute the number of dimension, one has to
not only select a volume-limited sample, but find
an area in the sky, where the catalog has been fully
completed. The 2dF catalog has two such areas, contains
about $70\,000$ points each. The 6dF catalog has
three such areas, each contains at least 20 000 points.
Identification of a volume-limited sample leaves
only about one third of all galaxies, implying that
significantly less than 10000 points should remain
in the areas of the 6dF catalog, which is evidently
insufficient for computing the fractal dimension. That
is why we calculate the number of dimension only for
the sample of 2dF galaxies.

We selected two sky areas almost completely covered
with observations:

\def\arraystretch{0}

\def\st{\raisebox{-1.2\dp\strutbox}{\rule{0pt}{1.8\ht\strutbox}}}

\begin{center}
\begin{tabular}{|c|c|c|c|}
\hline
\st \bf \bf Interval of $\alpha$ & \bf Interval of $\delta$ & $\Omega$ & \bf No. of galaxies\\

\hline \rule{0pt}{2pt} &&&\\ \hline

\st $150^\circ \div 210^\circ$ & $-4^\circ \div 2^\circ$ & $0.034\pi$ & 61259 \\

\hline

\st $328^\circ \div 52^\circ$ & $-32^\circ \div -24^\circ$ & $0.057\pi$ & 82044 \\

\hline

\end{tabular}
\end{center}

\begin{figure}\centering
\includegraphics{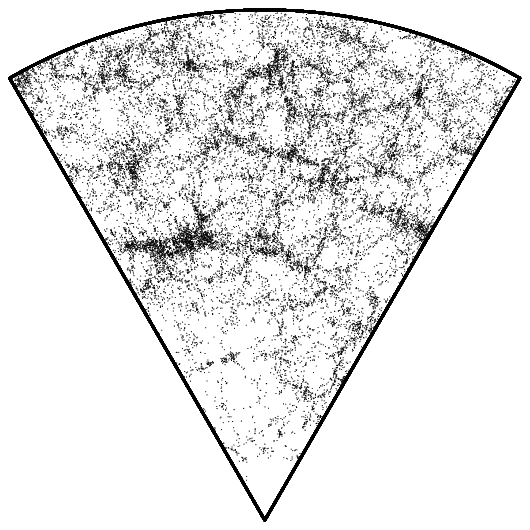}\\
\includegraphics{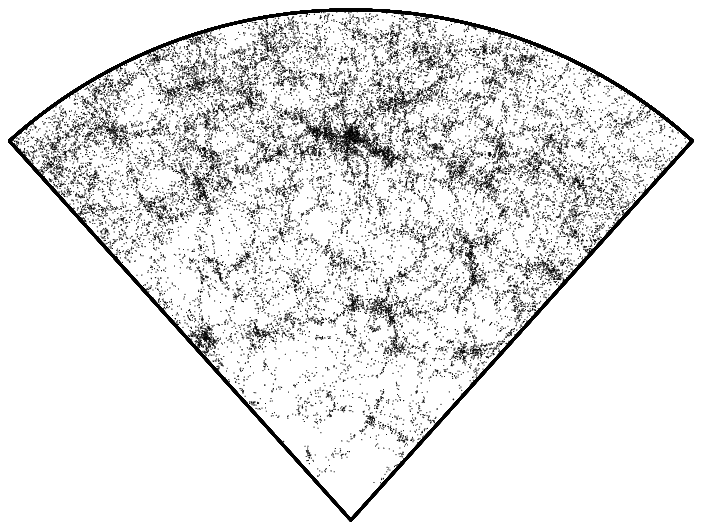}\\
\caption{Cone diagrams (in polar coordinates ($z,\alpha$))
of volume-limited samples up to $z_{lim}=0.15$ for the
Northern (top) and Southern (bottom) domains of the
2dF catalog.}\label{2dfcone}
\end{figure}

In each area we construct three volume-limited
samples: with redshifts of up to
$z=0.075$, $z=0.15$ and $z=0.2$, respectively. Figure \ref{2dfcone} shows the cone
diagrams for the volume-limited samples.

\subsection{Analysis of the Radial Distributions}
The analysis of radial distributions should not be
necessarily restricted to completely filled sky areas.
Therefore for both catalogs (2dF and 6dF) we construct
three radial distributions---one for the entire
set, and two for the two areas with sufficient observational
coverage. The corresponding areas for the 6dF
catalog are determined by the following parameters:
\begin{center}
\begin{tabular}{|c|c|c|c|}
\hline
\st \bf Interval of $\alpha$ & \bf Interval of $\delta$ & $\Omega$ & \bf No. of galaxies\\

\hline \rule{0pt}{2pt} &&&\\ \hline

\st $290^\circ \div 100^\circ$ & $-42^\circ \div -23^\circ$ & $0.263\pi$ & 16288 \\

\hline

\st $150^\circ \div 240^\circ$ & $-42^\circ \div -23^\circ$ & $0.139\pi$ & 14407 \\

\hline

\end{tabular}
\end{center}
For the 2dF catalog we select the same areas as those
used to determine the dimension.
\subsection{Conclusions}
\begin{figure}\centering
\includegraphics{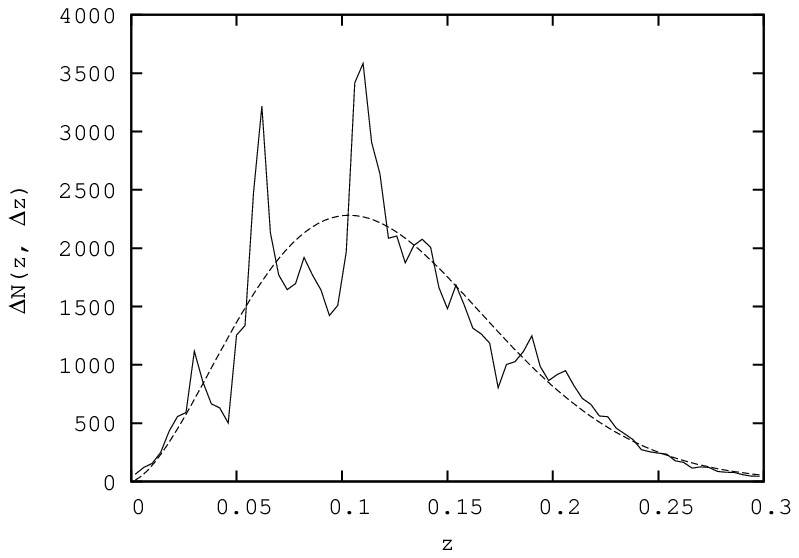}\\
\includegraphics{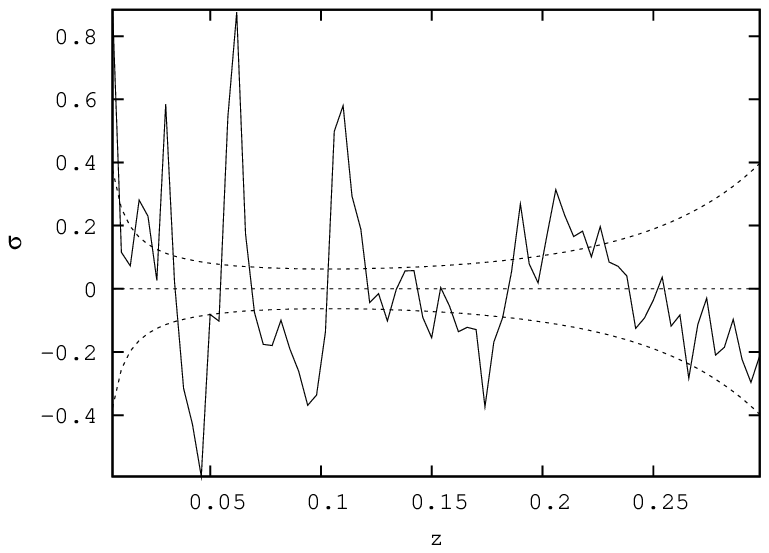}\\
\caption{Radial distribution (top) and the curve of relative
fluctuations (bottom) for the 2dF catalog. The first domain
$\Delta z = z_m/75$.}\label{dds1}
\end{figure}

\begin{figure}\centering
\includegraphics{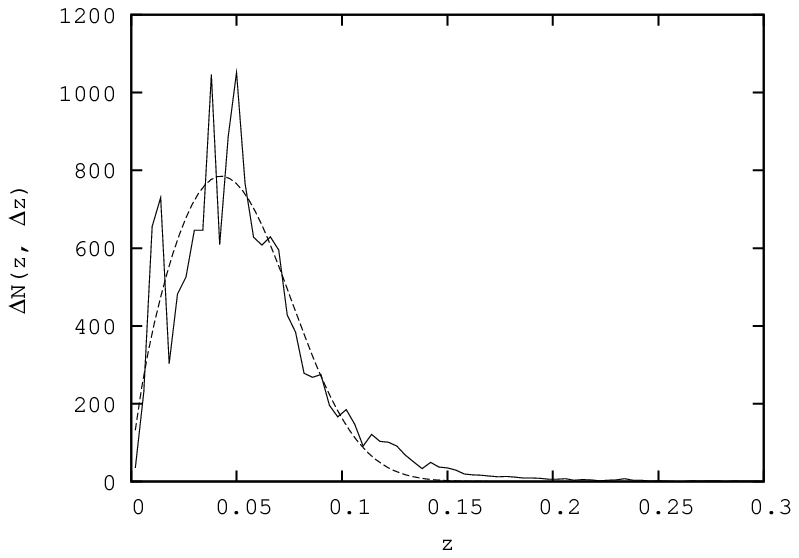}\\
\includegraphics{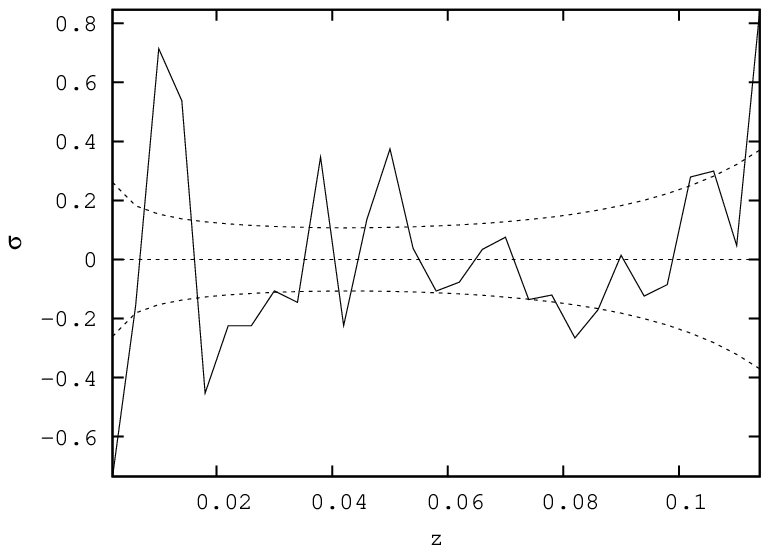}
\caption{Radial distribution (top) and the curve of relative
fluctuations (bottom) for the 6dF catalog. The first domain
$\Delta z = z_m/75$.}\label{dds2}
\end{figure}

\begin{itemize}
\item The result of computing of the number of dimensions
lead us to conclude that the fractal dimension
of the set of galaxies in volume-limited samples of
the 2dF catalog is equal to $2.20 \pm 0.25$. Irregularity
scales range from 2 Mpc to 20 Mpc (see Fig. \ref{2dfdim}). The
standard error of the dimension is greater than the
error for the simulated catalogs, what suggests that
the spatial distribution of galaxies is not ideally
fractal, but possibly multifractal.

We also tried to find the fractal dimension for the
6dF catalog in areas of the best observational coverage.
However, our analysis must have yielded grossly
underestimated dimension (1.5 and 1.9). This effect
is due to the insufficiently complete and insufficiently
uniform observational coverage of the sky area considered,
and different limiting redshifts $z$ in different
observational areas. Both of these effects produce
fictitious voids in the portion, where the dimension is
determined, resulting in its underestimated value.

\begin{figure}\centering
\includegraphics{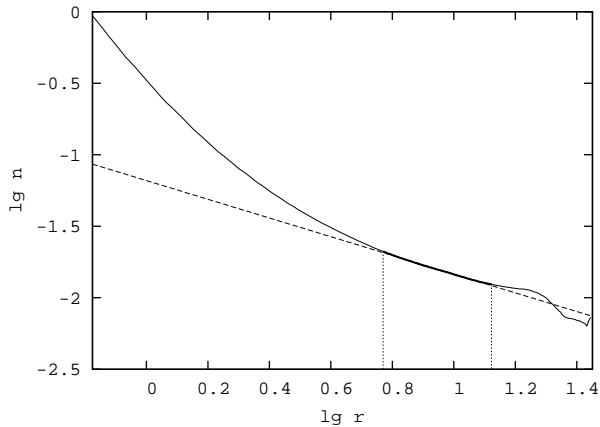}\\
\caption{Curve of conditional density for the 2dF catalog.
volume-limited sample in the first area of up to $z=0.075$.}\label{2dfdim}
\end{figure}

\item As for results of the analysis of radial distributions,
the main conclusion that follows from it consists in
the discovery of irregularities with the amplitudes
substantially exceeding not only $3\sigma_p$, but even
$7\sigma_p$ level, in the numbers much greater than one might
expect for a uniform distribution. This fact suggests
a clearly non-uniform distribution of galaxies up to
300--500 and 700 Mpc. Irregularities at the right
end of the radial distribution for the 6dF catalog
only indicate that the depth in $z$ varies for different observed areas and that the number of galaxies is
insufficient at large $z$. The characteristic sizes (scale
lengths) of irregularities amount to 40--70 Mpc (see
Fig. \ref{dds1}, \ref{dds2}).

The irregularity amplitude is of about 6, which
corresponds to a fractal dimension greater than 2.0,
but smaller than 2.6, i.e., the dimension estimate
$D = 2.2$, obtained using the correlation method, adequately
describes the nature of irregularities in the
radial distribution.
\end{itemize}

\section{CONCLUSIONS}
As a conclusion, we shall specify the following
results of our analysis of simulated catalogs of the
spatial distribution of galaxies, and absolute magnitude
distribution of galaxies:
\begin{itemize}
\item correlation methods can be correctly applied
only on scale lengths from several average distances
between the galaxies up to (10--20)\% of
the radius of the largest sphere that fits entirely
inside the set. The authors of earlier studies
believed that the method could be applied out
to the entire radius, and the above 10\% restriction,
which applies to all distributions but uniform,
was not taken into account when determining
the scale length where the distribution
becomes uniform (see, e.g., \cite{SL}, \cite{Tikhonov})
\item the empirical formula (\ref{efrr}), which is often used
to approximate radial distributions of objects
in magnitude-limited catalogs, yields
equally adequate minimum root-mean-square
approximation for both uniform and fractal
distributions with dimensions exceeding 2.0.
At smaller dimensions the scatter becomes too
large and the formula is inapplicable.

We found the fractal dimension to correlate
with the deviation of the true radial distribution
from the approximating formula, and not with
the parameters of the best-fit approximation.
\end{itemize}

Our analysis of real catalogs yielded the following
results:

\begin{itemize}
\item the data of the 2dF catalog imply a fractal dimension
of $2.20 \pm 0.25$ in the interval from 2 to
20 Mpc. No reliable conclusions can be made
on larger scales about the dimension and scale
of irregularities due to the intrinsic biases of
the method. Deeper surveys and surveys with
better sky coverage are needed for this task.

Because of its incompleteness, the 6dF catalog
can not yet be used to derive a reliable estimate
for the fractal dimension;

\item An analysis of radial distributions revealed
the significant irregularities both in the 2dF
and 6dF catalogs. Deviations from smooth
distribution exceed $7\sigma_p$ and their scale lengths
amount to 70 Mpc. The scale length and
magnitude of irregularities correlate rather well
with fractal-dimension estimates in the 2.1--2.4 interval.
\end{itemize}

\section*{ACKNOWLEDGMENTS}
I am sincerely grateful to Yu. V. Baryshev for formulating
the problem, for assistance and constant
attention to this work, and to V. P. Reshetnikov for his
useful advices and assistance in preparing the paper.
This work was supported by the Russian Foundation
for Basic Research (grant no. 09-02-00143).


\begin{thebibliography}{99}
\bibitem{Peebles} P. J. E. Peebles, astro-ph/0103040
\bibitem{Baryshev} Yu. V. Baryshev, P. Teerikorpi, Bull. Spec. Astrophys.
Obs. \textbf{59}, 92, 2006 
\bibitem{Vasiljev} N. V.Vasil'ev, MSc thesis (St.-Petersburg State University,
2004).
\bibitem{Erdogdu} P. Pirin Erdo\u gdu, O. Lahav, J. P. Huchra, M. Colless,
    et al., Monthly Notices Roy. Astronom. Soc. \textbf{373}, 45 (2006).
\bibitem{Massey} R. Massey, J. Rhodes, A. Leauthaud, et al., Astrophys.
J. Suppl. \textbf{172}, 239 (2007).
\bibitem{Somerville} R. S. Somerville, K. Lee, H. C. Ferguson, et al.,
           Astrophys. J. \textbf{600}, 171 (2004).
\bibitem{Busswell} G. S. Busswell, T. Shanks, W. J. Frith, et al., Monthly
Notices Roy. Astronom. Soc. \textbf{354}, 991 (2004).
\bibitem{Frith} W. J. Frith, G. S. Busswell, R. Fong, N. Metcalfe and
    T. Shanks, Monthly Notices Roy. Astronom. Soc. \textbf{345}, 1049 (2003).
\bibitem{Colles} M. Colless et al., astro-ph/0306581
\bibitem{Jones} D. H. Jones, W. Saunders, M. Colless, et al, Monthly
Notices Roy. Astronom. Soc. \textbf{355}, 747 (2004).
\bibitem{Jones2} D. H. Jones, W. Saunders, M. Read and M. Colless,
    PASA \textbf{22}, 277 (2005).
\bibitem{Wakamatsu} K. Wakamatsu, M. Colless, T. Jarrett, Q. Parker,
           W. Saunders and F. Watson, ASPC. \textbf{289}, 97 (2003).
\bibitem{MT} M. Matsumoto and T. Nishimura,
           ACM Transactions on Modeling and Computer Simulation,
           \textbf{8}, 1 (1998).
\bibitem{Felten} J. E. Felten, IAUS. \textbf{117}, 111 (1987).
\bibitem{Yoshii} Y. Yoshii and F. Takahara, Astrophys. J. \textbf{326}, 1 (1998).
\bibitem{Norberg} P. Norberg, S. Cole, C. M. Baugh, et al.,
           Monthly Notices Roy. Astronom. Soc. \textbf{336}, 907 (2002).
\bibitem{SL} F. Sylos Labini, M. Montuori and L. Pietronero,
           Phys. Rep. \textbf{293}, 61 (1998).
\bibitem{Tikhonov} A. V. Tikhonov et al., Bull. Spec. Astrophys.Obs. \textbf{50}, 39, 2000.        
\end{thebibliography}
\end{document}